\documentclass[9pt,twocolumn,twoside]{osajnl}

\journal{ol}
\setboolean{shortarticle}{true}

\usepackage{lineno}

\title{Encoding information in the mutual coherence of spatially separated light beams}

\author[1]{Alfonso~Nardi}
\author[1]{Shawn~Divitt}
\author[1]{Massimiliano~Rossi}
\author[1,2]{Felix~Tebbenjohanns}
\author[1]{Andrei~Militaru}
\author[1]{Martin~Frimmer}
\author[1,3,*]{Lukas~Novotny}

\affil[1]{Photonics Laboratory, ETH Zürich, Zürich, Switzerland}
\affil[2]{Currently with the Department of Physics, Humboldt-Universität zu Berlin, 10099 Berlin, Germany}
\affil[3]{Quantum Center, ETH Zürich, Zürich, Switzerland}

\affil[*]{Corresponding author: lnovotny@ethz.com}

\begin{abstract}
Coherence has been used as a resource for optical communications since its earliest days.
It is widely used for multiplexing of data, but not for encoding of data.
Here we introduce a coding scheme, which we call \textit{mutual coherence coding}, to encode information in the mutual coherence of spatially separated light beams.
We describe its implementation and analyze its performance by deriving the relevant figures of merit (signal-to-noise ratio, maximum bit-rate, and spectral efficiency) with respect to the number of transmitted beams. 
Mutual coherence coding yields a quadratic scaling of the number of transmitted signals with the number of employed light beams, which might have benefits for cryptography and data security.
\end{abstract}

\begin{document}

\maketitle

\textit{Introduction --}
Optical coherence is a fundamental degree of freedom of the light field.
It accounts for the statistical properties of the electromagnetic radiation~\cite{BornWolf}.
Not only has the study of optical coherence been crucial to the foundations and development of modern optics~\cite{wolf2007influence}, but it has also found countless applications in optical technology~\cite{Korotkova2020}.
In particular, a coherence-based multiplexing technique (known as \textit{coherence multiplexing}) has been proposed as the first coherent optical code-division multiple access (CDMA) technique~\cite{Sampson1997}.
Its appeal (in common with other CDMA techniques) lies in the fact that multiple users share the same optical bandwidth (relaxing the requirements for wavelength control) and transmit asynchronously to each other (providing independence of network synchronization).
Coherence multiplexing, in all the proposed configurations~\cite{Brooks1985, Griffin1995}, makes use of optical coherence to discriminate one pair of transmitted signal-reference beams from all the other pairs.
Although this CDMA technique uses coherence as a resource for multiplexing, its data encoding scheme  still relies on the relative optical path difference between two fields, requiring each signal to have a dedicated reference.
The dependence on a reference field can be eliminated by encoding information directly into the mutual coherence between pairs of transmitted light beams.
This scheme would not only eliminate the need of reference fields, but would also reduce the number of fields to be multiplexed.
In recent work we have introduced an approach to independently control the mutual coherence between pairs of spatially separated light beams~\cite{Nardi21}.
Here, we use a similar approach to implement a reference-less, coherence-based coding scheme, which we term {\em mutual coherence coding}.

By encoding information in the mutual coherence between pairs of spatially separated transmitted fields, we eliminate the need of a reference in the decoding step.
Furthermore, we gain a quadratic scaling of the number of transmitted signals with the number of transmitted light beams.
After introducing basic definitions we describe a general implementation that allows to encode and decode information in field-field correlations.
We discuss the dependence of relevant figures of merit (signal-to-noise ratio, maximum bit-rate and spectral efficiency) on the number of transmitted beams.
Finally, we discuss the benefits of using the proposed technique in combination with coherence multiplexing and generalizing the coherence control to other degrees of freedom.

\textit{Principle --}
\begin{figure*}
    \centering
    \includegraphics[width=0.75\textwidth]{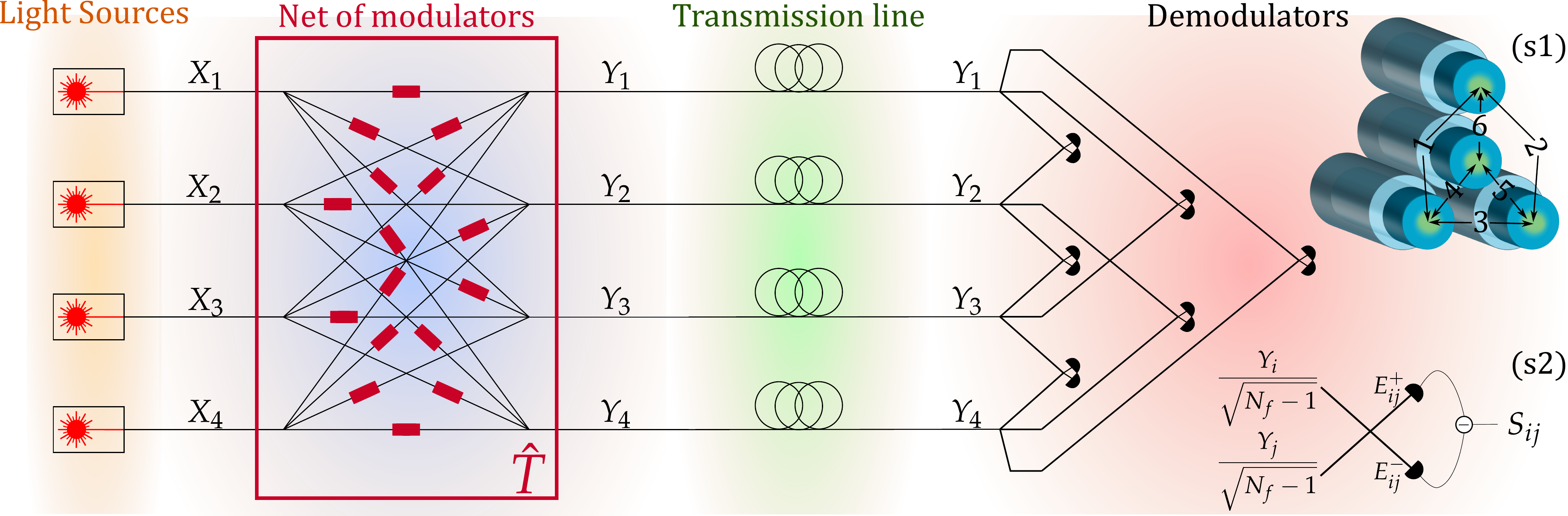}
    \caption{Implementation of mutual coherence coding. 
    A set of broadband light sources (here: 4) are fed into a linear port $\hat{T}$ to control their mutual coherences.
    Afterward, the fields reach the receiver through a transmission line.
    The received fields are then split to recover the encoded signals.
    The mutual coherence is reconstructed through pair-wise interference experiments, which can be implemented using a beam splitter and a balanced detector.
    Insets: (s1) sketch of four transmission fibers. The mutual coherence is defined for each pair, leading to a quadratic scaling of the number of transmitted signals;
    (s2) detection scheme. Two fields are mixed by a beam splitter and measured by a differential detector.
    }
    \label{fig:implementation}
\end{figure*}
The term {\em mutual coherence} quantifies the correlation between a pair of stochastic fields.
The mutual coherence can be defined with respect to any degree of freedom, e.g., space (spatial coherence), time (temporal coherence), polarization (degree of polarization~\cite{Wolf2007}), and transverse modes.
In this work we consider the special case of spatially separated optical beams.
We represent two random, statistically stationary, narrow-band electric fields at two different points in space by the complex analytic signals $E_i$ and $E_j$.
The mutual coherence (also known as \textit{equal-time degree of coherence}~\cite{BornWolf}) $\gamma_{ij}$ between $E_i$ and $E_j$ is defined as
\begin{equation}
\gamma_{ij}=\frac{ \overline{ E_{i}E_{j}^{\ast}} } { \sqrt{ \overline{ |E_{i}|^{2}} ~ \overline{ |E_{j}|^{2} } } } \;,
 \end{equation}
where the overline stands for ensemble or infinite time average (equivalent for ergodic and stationary fields).
The mutual coherence is a normalized quantity, spanning from full incoherence ($|\gamma_{ij}| = 0$) to partial ($0 < |\gamma_{ij}| < 1$) to full coherence ($|\gamma_{ij}| = 1$).
Let us now consider a set of $N_f$ mutually incoherent input fields $X_1, X_2, \dots, X_{N_f}$.
We use an optical linear port implementing the transformation $\hat{T}$ to get the output fields $Y_1, Y_2, \dots, Y_{N_f}$.
We define an output coherence matrix $\mathbb{K}$, such that the element $[\mathbb{K}]_{ij} = \gamma_{ij}$ is the mutual coherence between the fields $Y_i$ and $Y_j$.
The linear port can be chosen in such a way that we can program the desired output coherence matrix, as previously shown in Ref.~\cite{Nardi21}.
We note that $\mathbb{K}$ must be positive semi-definite, which is always fulfilled if $|\gamma_{ij}| \leq 1/(N_f-1)$, for all $i \neq j$.
This choice, while allowing us to independently choose the values of all the mutual coherences, restricts the analysis to a subspace of possible coherence matrices.
One could envision a more general scenario where the value of $\gamma_{ij}$ can exceed the mentioned bound (see Sec.~1 in Supplement 1).

Thanks to our ability to control the output coherence matrix, we can use the mutual coherence between pairs of output fields as our signal.
The transmission rate of a bit stream is then limited by the time needed to generate a desired coherence matrix, e.g., the modulators' refresh rate in Ref.~\cite{Nardi21}. 
Considering $N_f$ output fields, the number of signals equals the number of field pairs, that is $N_{p} = N_f(N_f - 1)/2$.
Thus, for large $N_f$ we achieve a quadratic scaling of the number of transmitted signals with the number of transmitted beams.

\textit{Implementation --}\label{sec:implementation}
A general implementation of mutual coherence coding is sketched in Fig.~\ref{fig:implementation}, for the case of $N_f = 4$.
The incoherent input fields $X_1, X_2, \dots, X_{N_f}$ have the same central frequency $\nu_0$, linewidth $\Delta\nu$ and power $P_\mathrm{in}$.
The light sources employed can be LEDs or laser diodes, characterized by a large linewidth, but can still be considered narrow-band, i.e., $\Delta\nu \ll \nu_0$.
%
%
Through a linear port, each input $X_j$ is connected to an output field $Y_i$, with $i \in \{1,2,\dots,N_f\}$, through a complex coefficient $t_{ij}$, i.e., $Y_i = \sum_{j=1}^{N_f} t_{ij} X_j$.
The mutual coherence $\gamma_{ij}$ between two output fields $Y_i$ and $Y_j$ takes the form
\begin{equation}\label{eq:gamma_from_Ys}
    \gamma_{ij} 
    = \frac{\overline{ Y_i Y_j^* }}{P_\mathrm{out}}
    = \sum_{n=1}^{N_f} t_{in} \sum_{m=1}^{N_f} t_{jm}^* \frac{\overline{ X_n X_m^* }}{P_\mathrm{out}}
    =  \frac{P_\mathrm{in}}{P_\mathrm{out}} \sum_{n=1}^{N_f} t_{in} t_{jn}^*\;,
\end{equation}
where we used the condition of input mutual incoherence $\overline{X_i X_j^*} = P_\mathrm{in} \delta_{ij}$, and we assumed the output fields having the same power $\overline{ | Y_i |^2 } = P_\mathrm{out} = \eta P_\mathrm{in}$, where $\eta = P_\mathrm{out}/P_\mathrm{in}$ is the transmission efficiency. 
%
%
By tuning the values of the modulation coefficients $t_{ij}$, we can then control the values of $\gamma_{ij}$~\cite{Nardi21}.
%
%
The linear port realizing the transformation $\hat{T}$ can be implemented in different ways, e.g., with free-space optics~\cite{Nardi21} or photonic integrated circuit~\cite{Miller13}.
The output fields $Y_i$ are transmitted to the receiver through a transmission line.
The transmission line in the case of signals carried by spatially separated light beams can be a bundle of single mode fibers, or a multi-core fiber, making use of the platform developed for spatial division multiplexing~\cite{Puttnam21}.
%
%
At the receiver side, we reconstruct the values of mutual coherences to decode the transmitted data.
We divide each received light beam into $N_f-1$ copies, and subsequently perform $N_f(N_f-1)/2$ pair-wise interference experiments.
%
%
The interference experiment is carried out by mixing two fields with a beam splitter, and measuring the intensities of the two output ports with a differential detector (see inset s2 of Fig.~\ref{fig:implementation}).
The two fields $E_{ij}^+$ and $E_{ij}^-$ at the output ports are
\begin{equation}
    E_{ij}^\pm = \frac{Y_i \pm Y_j}{\sqrt{2(N_f-1)}}\;,
\end{equation}
while the signal measured by the balanced detector is
\begin{equation}
    S_{ij} = \langle |E_{ij}^+|^2 \rangle  - \langle |E_{ij}^-|^2 \rangle = \frac{ 2~\mathrm{Re}\big( \langle Y_i Y_j^* \rangle \big) } { N_f-1 } \;.
\end{equation}
The angle brackets stand for the finite-time average performed by the detector, defined as 
\begin{equation}
	\langle Y_i Y_j^* \rangle = \frac{1}{T} \int_{-T/2}^{T/2} Y_i Y_j^* dt \; .
\end{equation} 
The detected signal is a random process, which we can characterize with an expectation value and a variance.
From the expectation value, we confirm that the designed measurement corresponds indeed to an unbiased estimate of the real value of the mutual coherence
\begin{equation}
    \overline{ S_{ij} } = 
    \frac{2}{T(N_f-1)} \int_{-T/2}^{T/2}~ \mathrm{Re}\left( \overline{ Y_i Y_j^* } \right) dt = \frac{2 P_\mathrm{out}}{N_f-1} \mathrm{Re}(\gamma_{ij}) \;,
\end{equation}
where we used Eq.~(\ref{eq:gamma_from_Ys}) for the definition of $\gamma_{ij}$.
However, the stochastic nature of the underlying fields leads to a non-zero variance even in the case where no noise is considered (see Sec.~2 in Supplement 1 for the derivation)
\begin{equation}\label{eq:optical_beat_noise}
    \text{Var}( S_{ij} ) 
    = \overline{ S_{ij}^2 } - \left(\overline{ S_{ij} }\right)^2 
    = \frac{2 P_\mathrm{out}^2 \left[1 + \mathrm{Re}\left( \gamma_{ij}^2 \right)\right]}{(N_f-1)^2 T\Delta\nu} \;.
\end{equation}
This variance is inversely proportional to the number of statistically independent realizations that we sample during the measurement time $T$.
Fields separated by a time longer than the inverse of the linewidth are uncorrelated, hence the number of independent realizations of the random process is given by $\Delta \nu T$~\cite{Loudon2000}.
Moreover, in the limit of small mutual coherence $|\gamma_{ij}| \ll 1$, the variance saturates to a non-zero value determined by the number of collected samples.
The variance in Eq.~(\ref{eq:optical_beat_noise}) gives a contribution to the total noise, which is known in the literature as \textit{optical beat noise}, and it is regarded as the main limitation for the performance of traditional coherence multiplexing~\cite{Pendock1997}.

\textit{Performance --}\label{sec:performances}
\begin{figure}
    \centering
    \includegraphics[width=0.9\columnwidth]{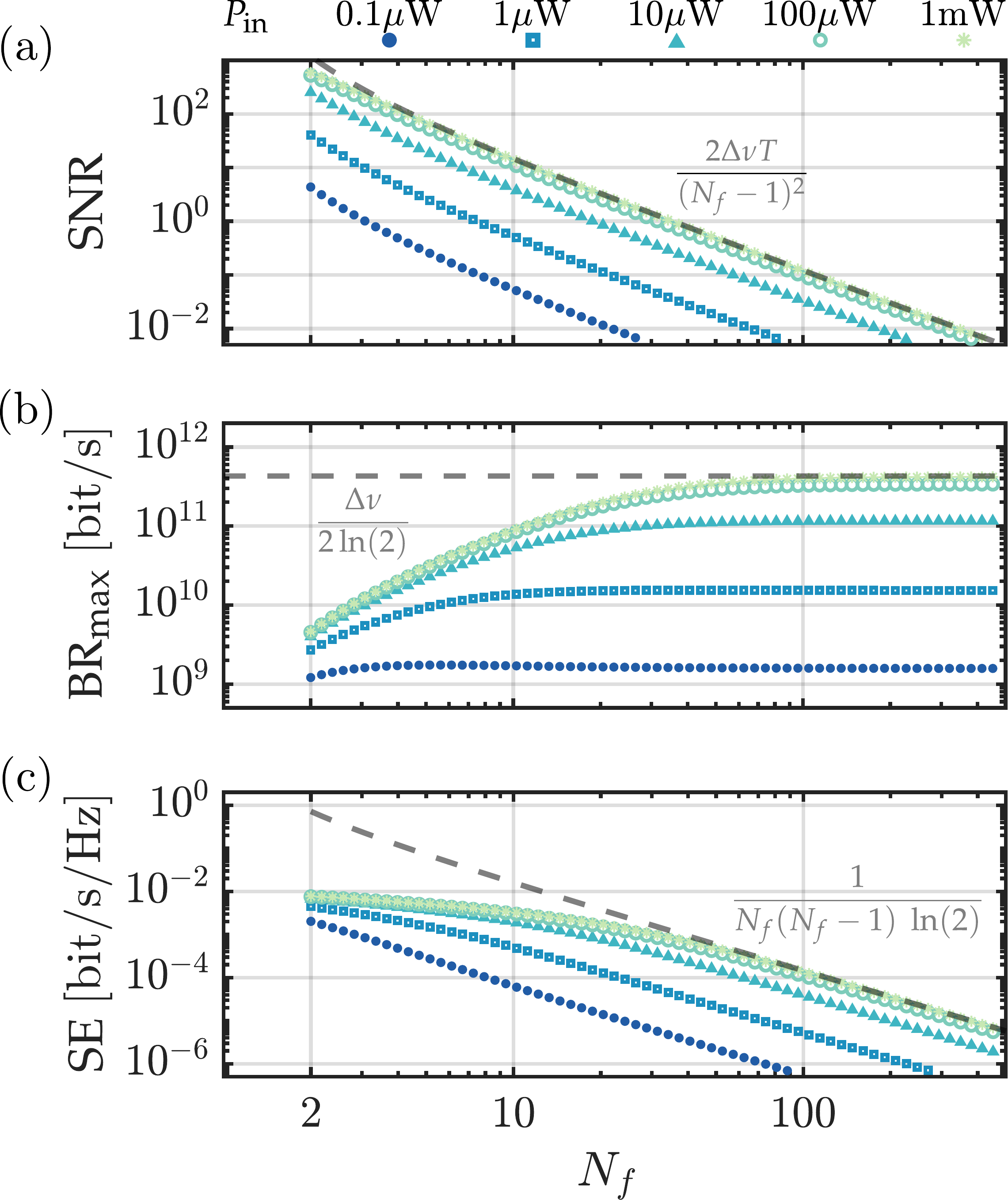}
    \caption{System performance.
    (a) $\mathrm{SNR}$, (b) maximum bit-rate ($\mathrm{BR_{max}}$) and (c) spectral efficiency ($\mathrm{SE}$) as a function of the number of transmitted fields ($N_f$) and input power ($P_\mathrm{in}$).
    The dashed lines correspond to (a) Eq.~\ref{eq:limit_SNR}, (b) Eq.~\ref{eq:limit_BR} and (c) Eq.~\ref{eq:expression_SE}.
    The system parameters are: $T = 1~\mathrm{ns}$, $\gamma_\mathrm{max} = 1/(N_f-1)$, $\nu_0 = 870~\mathrm{nm}$, $\Delta\nu = 1~\mathrm{nm}$ and $\eta = 1\%$.
    }
    \label{fig:FiguresOfMerit}
\end{figure}
We will now discuss important figures of merit for assessing the performance of mutual coherence coding, starting with the bit-rate.
For the sake of generality, we choose as a reference the maximum bit-rate $\mathrm{BR_{max}}$ allowed by the channel capacity in the presence of additive white gaussian noise~\cite{Shannon1949}. 
The mutual coherence between each field pair is an independent signal, hence the number of independent communication channels present in the system is equal to the number of field pairs $N_p$, leading to a maximum bit-rate
\begin{equation}\label{eq:BRmax_def}
    \mathrm{BR_{max}} = N_{p} ~ \frac{ \log_2\left(1 + \mathrm{SNR}\right) } {2T} \;.
\end{equation}
The signal-to-noise ratio ($\mathrm{SNR}$) is defined as
\begin{equation}
    \mathrm{SNR} = \frac{ \left( \overline{ \tilde S_{ij} } \right)^2 } { \text{Var} ( \tilde S_{ij}  ) } \;,
\end{equation}
where $\tilde S_{ij} $ is the detected signal resulting from the mutual coherence of the pair of fields $Y_i$ and $Y_j$ in the presence of noise.
For multiple, independent noise sources the variance is the sum of the variances of the different noise contributions (see Sec.~3 in Supplement 1).
Restricting ourselves to optical beat noise and shot noise, the $\mathrm{SNR}$ can be expressed as
\begin{equation} \label{eq:SNR_expression}
    \mathrm{SNR} = \left[ \frac{ 1 + \mathrm{Re}(\gamma_{ij}^2) } { 2 \mathrm{Re}(\gamma_{ij})^2 ~ \Delta\nu T } + \frac{ h \nu_0 } { \eta P_\mathrm{in} \mathrm{Re}(\gamma_{ij})^2 ~ T } \right]^{-1} \;.
\end{equation}
The first term is the optical beat noise highlighted in \eqref{eq:optical_beat_noise} and the second term corresponds to shot noise.
The derivation of Eq.~(\ref{eq:BRmax_def}) assumes a value of $\mathrm{Re}(\gamma_{ij}) = \gamma_\mathrm{max}$ that is at the extreme of the allowed range of values. 
In fact, with this choice the $\mathrm{SNR}$ provides the maximum number of discretization steps of the signal. 
We choose $\gamma_\mathrm{max} = 1/(N_f - 1)$ to ensure the positive semi-definiteness of the output coherence matrix.
We then insert $\gamma_\mathrm{max}$ into Eq.~(\ref{eq:SNR_expression}) for the $\mathrm{SNR}$ and then into Eq.~(\ref{eq:BRmax_def}) to obtain the maximum bit-rate.
In Figs.~\ref{fig:FiguresOfMerit}a and \ref{fig:FiguresOfMerit}b, we show the $\mathrm{SNR}$ and the $\mathrm{BR_{max}}$, respectively, as a function of the number of transmitted fields $N_f$ for different input powers.
The $\mathrm{SNR}$ decays monotonically with increasing $N_f$, whereas the maximum bit-rate first increases quadratically and then saturates at large $N_f$ values.
The quadratic increase derives from the number of field pairs, while the saturation is due to the low $\mathrm{SNR}$.
In the limit of large $N_f$ and sufficient input power (to be limited by optical beat noise) the $\mathrm{SNR}$ takes the expression
\begin{equation}\label{eq:limit_SNR}
    \mathrm{SNR} \approx \frac{2 \Delta \nu T}{(N_f-1)^2} \;,
\end{equation}
where the factor $\Delta \nu T$ can be understood as the number of independent realizations of the light field within the time $T$.
When the $\mathrm{SNR}$ becomes smaller than $1$, we can Taylor expand Eq.~(\ref{eq:BRmax_def}) and the maximum bit-rate saturates at the value
\begin{equation} \label{eq:limit_BR}
    \mathrm{BR_{max}} \approx \frac{N_f(N_f-1)}{4T} \left[ \frac{2 \Delta\nu T}{(N_f-1)^2 \ln(2)} \right] \approx \frac{\Delta\nu}{2\ln(2)} \;.
\end{equation}
Therefore, the ultimate limit of the bit-rate is given by the input linewidth.
The saturation can be well understood considering the statistical nature of mutual coherence, since the rate of field realizations needed to reconstruct the mutual coherence is given by the linewidth.


Alongside the maximum bit-rate, an important figure of merit is the spectral efficiency (SE), which quantifies how well the available bandwidth is used.
It is defined as the bit-rate per communication channel per unit bandwidth $(\mathrm{bit/s/Hz})$, i.e.,
\begin{equation}\label{eq:expression_SE}
    \mathrm{SE} = \frac{\mathrm{BR_{max}}}{N_p~\Delta\nu} < \frac{1}{N_p~2\ln(2)} = \frac{1  } { N_f (N_f-1)~\ln(2) }\;.
\end{equation}
The dependence of the $\mathrm{SE}$ on the number of transmitted fields $N_f$ is represented in Fig.~\ref{fig:FiguresOfMerit}c.
With a value much smaller than one, mutual coherence coding performs worse in terms of spectral efficiency than standard coding techniques~\cite{Yasaka2017}.
On the other hand, the bandwidth $\Delta\nu$ of mutual coherence coding is considerably larger than the bandwidth of other coding techniques, which are restricted by the bandwidth of optoelectronic components, such as detectors and modulators ($\sim 50~\mathrm{GHz}$).
The large bandwidth associated with mutual coherence coding is provided by the input sources, and it is used to build the statistical ensemble required to achieve a quadratic scaling of the number of transmitted signals on $N_f$.
Thanks to this quadratic scaling, mutual coherence coding can achieve bit-rates on the order of $\Delta \nu$, hence very high for input with large linewidth.
Therefore, even if the spectral efficiency of mutual coherence coding is much lower than in other coding schemes, we can gain technical advantage in terms of bit-rate.

\textit{Discussion --}
Mutual coherence coding comes with a quadratic scaling of the bit-rate with the number of transmitted fields, up to the point where it saturates.
The saturation value is given by the linewidth of the light sources employed.
Selecting broadband light sources, such as LEDs or laser diodes, can therefore lead to remarkable data transmission rates.
These light sources have the added advantage of being inexpensive, thus facilitating the scalability.
Furthermore, the use of broadband light makes our proposal suitable for free-space communications, where partially coherent sources have been reported to be more robust to atmospheric turbulence~\cite{Korotkova2004}.
Another advantage of the proposed method is the absence of local oscillators at the receiver side, since in mutual coherence coding the transmitted beams are mutually referenced, allowing for a simpler implementation.
However, the discussed implementation requires an accurate control on the length difference between the transmitted fibers.
In fact, through the linear port we can control the \textit{equal-time} mutual coherence, which is conserved only as long as the optical path difference of the interfering beams does not exceed the coherence length of the input source.
This issue becomes more severe for large input linewidths, as the coherence length is reduced accordingly.
A possible solution to this limitation is to combine mutual coherence coding and coherence multiplexing, obtaining a fully coherence-based communication system.
To do so, one would introduce, at the sender side, a different delay for each output beam of the linear port.
If the individual delays and their differences are much longer than the coherence time of the light sources, the fields will not interfere, even when combined in the same single mode fiber.
The receiver then splits the transmitted field and compensates for the input delays to reconstruct the multiplexed signals, before proceeding with the pair-wise interference experiments.
Besides allowing to perform the transmission with a single fiber, this approach relaxes the requirement on the length difference between the fibers.
In fact, since the transmission line is common, any optical path difference between the interfering beams can only originate from the introduced delays.
Interestingly, in this approach the delay lines are very short (on the order of the coherence length) and embedded in a well controlled environment.
The needed calibration of the delay length can then make use of the techniques developed for standard coherent detection schemes~\cite{Wu2021}.

Moreover, having a coherence based communication system comes with security benefits.
Suppose the sender and the receiver have a pre-shared set of delays (which can be frequently updated).
An eavesdropper cannot record the message encoded in the mutual coherence, because they are not able to record the fluctuations of the light field in real time.
An eavesdropper could measure the mutual coherences only through interference experiments and only if they knew the pre-shared delays.
This security benefit is similar to that of traditional coherence multiplexing~\cite{Wacogne1996}.

As a concluding remark, we highlight that mutual coherence coding is not limited to spatial coherence.
Coherence based on any other degree of freedom of light (temporal coherence, degree of polarization, correlations between transverse modes, etc.) can be used to encode information, allowing for various implementations.
Moreover, resorting to mutual coherence between different degrees of freedom~\cite{Kagalwala2015} can lead to a great enhancement of the scaling with the number of transmitted fields.
Considering for example $n$ different beams (spatial degree of freedom), each of them characterized by $m$ transverse modes, the number of mutual coherences that we can control is $N_{p} \approx (n \times m)^2/2$.
This can be further generalized (e.g., including time and polarization) to greatly improve the capacity of current communication systems. 


\textit{Conclusions --}
In this work, we proposed a new method to encode information in the mutual coherence of separate light beams.
Light coherence is not only used as a resource for multiplexing different signals in the same medium, but also for encoding data.
We derived the $\mathrm{SNR}$, the maximum bit-rate and the spectral efficiency with respect to the number of transmitted light beams.
We discussed the benefits of employing mutual coherence coding, in particular the potentially high bit-rate and the added security resulting from the combination with coherence multiplexing.
Finally, we highlighted that mutual coherence coding can be generalized by involving other degrees of freedom of light, which could lead to an even larger scaling of number of transmitted signals per transmitted field.

\paragraph{Funding.} ETH Zurich Research Grant (ETH-41 19-1).

\paragraph{Disclosures.} The authors declare that there are no conflicts of interest related to this article.

\paragraph{Data availability.} No data were generated or analyzed in the presented research.

\paragraph{Supplemental document.} See Supplement 1 for supporting content.

\bibliography{references}

\begin{thebibliography}{10}
\newcommand{\enquote}[1]{``#1''}

\bibitem{BornWolf}
M.~Born and E.~Wolf, \emph{Principles of Optics} (Cambridge University, 1999),
  7th ed.

\bibitem{wolf2007influence}
E.~Wolf, {\protect\JournalTitle{Progress in Optics}} \textbf{50}, 251 (2007).

\bibitem{Korotkova2020}
O.~Korotkova and G.~Gbur, {\protect\JournalTitle{Progress in Optics}}
  \textbf{65}, 43 (2020).

\bibitem{Sampson1997}
D.~D. Sampson, G.~J. Pendock, and R.~A. Griffin, {\protect\JournalTitle{Fiber
  and Integrated Optics}} \textbf{16}, 129 (1997).

\bibitem{Brooks1985}
J.~Brooks, R.~Wentworth, R.~Youngquist, M.~Tur, B.~Kim, and H.~Shaw,
  {\protect\JournalTitle{Journal of Lightwave Technology}} \textbf{3}, 1062
  (1985).

\bibitem{Griffin1995}
R.~Griffin, D.~Sampson, and D.~Jackson, {\protect\JournalTitle{Journal of
  Lightwave Technology}} \textbf{13}, 1826 (1995).

\bibitem{Nardi21}
A.~Nardi, F.~Tebbenjohanns, M.~Rossi, S.~Divitt, A.~Norrman, S.~Gigan,
  M.~Frimmer, and L.~Novotny, {\protect\JournalTitle{Opt. Express}}
  \textbf{29}, 40831 (2021).

\bibitem{Wolf2007}
E.~Wolf, \emph{Introduction to the Theory of Coherence and Polarization of
  Light} (Cambridge University, 2007).

\bibitem{Miller13}
D.~A.~B. Miller, {\protect\JournalTitle{Photon. Res.}} \textbf{1}, 1 (2013).

\bibitem{Puttnam21}
B.~J. Puttnam, G.~Rademacher, and R.~S. Lu\'{i}s,
  {\protect\JournalTitle{Optica}} \textbf{8}, 1186 (2021).

\bibitem{Loudon2000}
R.~Loudon, \emph{The Quantum Theory Of Light} (Oxford University, 2000), 3rd
  ed.

\bibitem{Pendock1997}
G.~J. Pendock and D.~D. Sampson, {\protect\JournalTitle{Optics Communications}}
  \textbf{143}, 109 (1997).

\bibitem{Shannon1949}
C.~Shannon, {\protect\JournalTitle{Proceedings of the IRE}} \textbf{37}, 10
  (1949).

\bibitem{Yasaka2017}
H.~Yasaka and Y.~Shibata, \emph{Fibre Optic Communication: Key Devices}
  (Springer International Publishing, Cham, 2017), chap. Semiconductor-Based
  Modulators, pp. 359--416.

\bibitem{Korotkova2004}
O.~Korotkova, L.~C. Andrews, and R.~L. Phillips, {\protect\JournalTitle{Optical
  Engineering}} \textbf{43}, 330  (2004).

\bibitem{Wu2021}
R.~Wu, F.~Yang, Y.~Sun, N.~Cheng, J.~Wang, F.~Wei, Y.~Gui, and H.~Cai,
  {\protect\JournalTitle{Opt. Express}} \textbf{29}, 14041 (2021).

\bibitem{Wacogne1996}
B.~Wacogne and D.~Jackson, {\protect\JournalTitle{IEEE Photonics Technology
  Letters}} \textbf{8}, 947 (1996).

\bibitem{Kagalwala2015}
K.~H. Kagalwala, H.~E. Kondakci, A.~F. Abouraddy, and B.~E.~A. Saleh,
  {\protect\JournalTitle{Scientific Reports}} \textbf{5}, 15333 (2015).

\end{thebibliography}


\begin{thebibliography}{1}
\newcommand{\enquote}[1]{``#1''}

\bibitem{supp:Devroye1986}
L.~Devroye, \emph{Non-Uniform Random Variate Generation}, (Springer New York, New York, NY, 1986).

\bibitem{supp:Wolkowicz1980}
H.~Wolkowicz and G.~P.~Styan, \enquote{Bounds for eigenvalues using traces,} Linear Algebr. its
Appl. 29, 471–506 (1980).

\bibitem{supp:Loudon2006} 
R.~Loudon, \emph{The quantum theory of light}, Oxford science publications (Oxford Univ. Press,
Oxford, 2006), 3rd ed.

\end{thebibliography}

\setcounter{figure}{0}
\setcounter{equation}{0}
\setcounter{section}{0}
\renewcommand{\thefigure}{S\arabic{figure}}
\renewcommand{\theequation}{S\arabic{equation}} 
\onecolumn

\begin{center}
	\textbf{\LARGE Encoding information in the mutual coherence of spatially separated light beams: \\supplemental document} \\ 
\end{center}
\vspace{3mm}

\author{} 

\begin{abstract}
\end{abstract}

\setboolean{displaycopyright}{false} 

\maketitle


\section{Coherence matrix properties}
Given a set of light fields, the coherence matrix $\mathbb{K}$ contains all the information about the coherence in the system. 
The diagonal elements of the matrix are the self degrees of coherence $\gamma_{ii}$, which are always equal to $1$, and the off-diagonal terms are the values of mutual coherence $[\mathbb{K}_{ij}] = \gamma_{ij}$. 
The coherence matrix is also known as a statistical correlation matrix, which is a normalized covariance matrix, and must be Hermitian and positive semi-definite \cite{supp:Devroye1986}.
From the Hermitian condition we have $\gamma_{ji} = \gamma_{ij}^*$, while the positive semi-definite condition poses that \cite{supp:Wolkowicz1980}
\begin{equation}
    \frac{tr(\mathbb{K})^2}{tr(\mathbb{K}^2)} \geq N_f - 1\;,
\end{equation}
where $tr$ is the matrix trace, and $N_f$ is the dimension of $\mathbb{K}$.
Combining the previous properties, we get a condition on the absolute values of the mutual coherences
\begin{equation}\label{eq:coeff_cond_pos_semdef}
    \sum_{i=1}^{N_f} \sum_{j=1}^{N_f} |\gamma_{ij}|^2 \leq \frac{N_f^2}{N_f-1} \;.
\end{equation}
Choosing all mutual coherences with the same absolute value, i.e., $|\gamma_{ij}|=|\gamma|$ $\forall \; i,j$, Eq.~(\ref{eq:coeff_cond_pos_semdef}) becomes
\begin{equation}\label{eq:strict_condition}
    |\gamma| \leq \frac{1}{N_f-1}\;.
\end{equation}
Choosing $|\gamma_{ij}|<1/(N_f-1)$ $\forall \; i,j$ would directly satisfy the condition in Eq.~(\ref{eq:strict_condition}), ensuring that the resulting coherence matrix is positive semi-definite, regardless of the phase of each mutual coherence $\gamma_{ij}$.
A more sophisticated choice [based on Eq.~(\ref{eq:coeff_cond_pos_semdef})] could lead to better performance of mutual coherence coding.


\section{Detected signal variance without added noise} \label{app:variance_derivation}
In this appendix, we derive the expression of the variance of the detected signal in mutual coherence coding.
We highlight that the variance considered here results from the stochastic nature of the mutual coherence: no noise is added to the transmitted fields in this treatment.
Let us consider the interference experiment between a single pair of fields $Y_i$ and $Y_j$.
The signal $S_{ij}$ measured by the balanced photodetector is (see main text)
\begin{equation}
    S_{ij} = \frac{2}{T(N_f-1)} \int_{-T/2}^{T/2}~ \mathrm{Re}\left(  Y_i Y_j^*  \right) dt  \;,
\end{equation}
where $T$ is the integration time of the detector, and $N_f$ is the number of transmitted fields.
The fields are stochastic variables, hence we characterize the signal through its expectation value and its variance.
The expectation value, denoted with an overline, is
\begin{equation}\label{eq:exp_value}
    \overline{ S_{ij} } = 
    \frac{2}{T(N_f-1)} \int_{-T/2}^{T/2}~ \mathrm{Re}\left( \overline{ Y_i Y_j^* } \right) dt = \frac{2 P_\mathrm{out}}{N_f-1} \mathrm{Re}(\gamma_{ij}) \;,
\end{equation}
where we used the relation $P_\mathrm{out} \gamma_{ij} = \overline{Y_i Y_j^*}$.
To derive the variance $\mathrm{Var}(S_{ij}) = \overline{S_{ij}^2} - \left(\overline{S_{ij}}\right)^2$, we compute the second moment
\begin{equation}
\begin{aligned}
    \overline{S_{ij}^2} 
    &= \frac{1}{T^2({N_f}-1)^2} \iint_{-T/2}^{T/2} ~ \overline{ \left[Y_i(t_1) Y_j^*(t_1) + c.c.\right]\left[Y_i(t_2) Y_j^*(t_2) + c.c.\right] } ~dt_1  dt_2 \\
    &= \frac{2}{T^2({N_f}-1)^2}  \iint_{-T/2}^{T/2} ~ \mathrm{Re} \bigg\{ \overbrace{ \overline{ \left[Y_i(t_1) Y_j^*(t_1) + c.c.\right] Y_i(t_2) Y_j^*(t_2) } }^{\mathcal{I}} \bigg\} ~dt_1  dt_2\;,
\end{aligned}
\end{equation}
where $c.c.$ stands for complex conjugate.
Recalling that the input fields $X_n$, with $n \in \{1,2,\dots,N_f\}$, are related to the outputs of the linear port through the relation $Y_i = \sum_n^{N_f} t_{in} X_n$, we can express $\mathcal{I}$ as:
\begin{equation}\label{eq:encountering_second_order}
\begin{aligned}
    \mathcal{I} 
    &=\overline{ \left[ Y_i(t_1)  Y_j^*(t_1) +  Y_i^*(t_1)  Y_j(t_1)\right]  Y_i(t_2)  Y_j^*(t_2)  } \\
    &= \sum_{n,m,p,q}^{N_f}  \left(t_{jn}^* t_{im} + t_{in}^* t_{jm} \right)  t_{jp}^* t_{iq} \overline{ X^*_n(t_1) X_m(t_1) X^*_p(t_2) X_q(t_2) } \;.
\end{aligned}
\end{equation}
To simplify this equation, we need to derive the expression for the second-order correlation function $\Gamma$ of the input fields, i.e.,
\begin{equation}\label{eq:definitionGamma}
    \Gamma_{nmpq}(t_1-t_2) = \overline{ X^*_n(t_1) X_m(t_1) X^*_p(t_2) X_q(t_2) }\;.
\end{equation}
The input fields are independent and have a null expectation value, hence $\Gamma_{nmpq}(t_1-t_2)$ is non-zero only if the values of the indexes are all equal, or equal in pairs. 
Therefore, only for the following four, mutually exclusive cases, the expression does not vanish. 
Firstly, if $n=m\neq p=q$, we have
\begin{equation}
    \Gamma_{nnpp}(t_1-t_2) = \overline{ |X_n(t_1)|^2 } \; \overline{ |X_p(t_2)|^2 } = P_\mathrm{in}^2.
\end{equation}
Secondly, if $n=q\neq m=p$, we have
\begin{equation}
    \Gamma_{nmmn}(t_1-t_2) = \overline{ X^*_n(t_1) X_n(t_2) } \; \overline{ X_m(t_1) X^*_m(t_2) } \;.
\end{equation}
To solve this equation, we need to introduce the degree of first-order coherence $g^{(1)}$ \cite{supp:Loudon2006}:
\begin{equation}
	g^{(1)}(t_1 - t_2) = \frac{\overline{ X^*_n(t_1) X_n(t_2) }}{P_\mathrm{in}}\;.
\end{equation}
We can consider the process stationary, hence not dependent on the particular value of $t_1$ and $t_2$, but only on their difference $\tau = t_1 - t_2$.
Moreover, for our input sources (which we consider chaotic) the expression of $g^{(1)}(\tau)$ is~\cite{supp:Loudon2006}
\begin{equation}\label{eq:g1_definition}
	g^{(1)}(\tau) = \exp \Big( - i 2 \pi \nu_0 \tau - |\tau| \Delta\nu \Big)\;,
\end{equation}
where $\nu_0$ and $\Delta\nu$ are the central frequency and the linewidth of the input light fields, respectively.
Thus, $\Gamma_{nmmn}$ takes the form
\begin{equation}
\Gamma_{nmmn}(\tau) = P_\mathrm{in}^2 \left| g^{(1)}(\tau) \right|^2  \;.
\end{equation}
Thirdly, if $n=p \neq m=q$, we have
\begin{equation}
    \Gamma_{npnp}(t_1-t_2) = \overline{ X^*_n(t_1) X^*_n(t_2) } \; \overline{ X_p(t_1) X_p(t_2) } = 0\;.
\end{equation}
Finally, we have for $n = m = p = q$
\begin{equation}
    \Gamma_{nnnn}(t_1-t_2) = \overline{ X^*_n(t_1) X_n(t_1) X_n^*(t_2) X_n(t_2) } = P_\mathrm{in}^2 g^{(2)}(\tau),
\end{equation}
where we introduced the degree of second-order coherence $g^{(2)}(\tau)$. 
With classical chaotic light sources we have \cite{supp:Loudon2006}
\begin{equation}
    g^{(2)}(\tau) = 1 + |g^{(1)}(\tau)|^2\;,
\end{equation}
which allows us to join the four cases into a single expression:
\begin{equation}\label{eq:second_moment_Gamma_to_g1}
    \Gamma_{nmpq}(\tau) = P_\mathrm{in}^2 \left[\delta_{nm}\delta_{pq} + \delta_{nq}\delta_{mp} |g^{(1)}(\tau)|^2 \right]\;.
\end{equation}
Going back to the integral $\mathcal{I}$ of Eq.~(\ref{eq:encountering_second_order}) we get
\begin{equation}
    \mathcal{I} = P_\mathrm{in}^2 \sum_{n,p}^{N_f}  \left[ \left(t_{jn}^* t_{in} + t_{in}^* t_{jn} \right) t_{jp}^* t_{ip}  + \left(t_{jn}^* t_{ip}  + t_{in}^* t_{jp} \right) t_{jp}^* t_{in} |g^{(1)}(t_1-t_2)|^2 \right] \;.
\end{equation}
From our choice of the linear port, we have $P_\mathrm{out} \gamma_{ij} = P_\mathrm{in} \sum_{n}^{N_f} t_{in} t_{jn}^*$ [see Eq.~(2) in the main text], leading to
\begin{equation}
    \mathcal{I} = P_\mathrm{out}^2 \left[ \left(\gamma_{ij} + \gamma_{ij}^* \right) \gamma_{ij} + \left( \gamma_{ij}^2 + \gamma_{ii}\gamma_{jj} \right) |g^{(1)}(\tau)|^2 \right] \;.
\end{equation}
Including the derived expression of $\mathcal{I}$ in the second moment of the signal and recalling that $\gamma_{ii} = 1$ we get
\begin{equation}\label{eq:det_second_moment}
    \overline{S_{ij}^2} = 
    \frac{ 4 P_\mathrm{out}^2 \mathrm{Re}(\gamma_{ij})^2 } { ( N_f - 1 )^2 } + \frac{ 2 P_\mathrm{out}^2 \left[ 1 + \mathrm{Re}(\gamma_{ij}^2) \right] } { T ( N_f - 1 )^2 }  \int_{-T/2}^{T/2} ~ |g^{(1)}(\tau)|^2 ~d\tau\;.
\end{equation}
The integral can be analytically solved from the definition of $g^{(1)}(\tau)$ given in Eq.~(\ref{eq:g1_definition}):
\begin{equation}
    \int_{-T/2}^{T/2} ~ |g^{(1)}(\tau)|^2 ~d\tau = \frac{1-\exp(-T\Delta\nu)}{\Delta\nu} \to \frac{1}{\Delta\nu}
\end{equation}
where we considered the limit of integration time $T$ much longer than the coherence time $\tau_c=1/\Delta\nu$ of the light sources.
Finally, the variance of the signal is
\begin{equation}
    \text{Var}( S_{ij} ) = \frac{2 P_\mathrm{out}^2 \left[1 + \mathrm{Re}\left( \gamma_{ij}^2 \right)\right]}{(N_f-1)^2 T\Delta\nu} \;.
\end{equation}


\newcommand{\BW}{\mathrm{BW}}
\section{Signal-to-noise ratio (SNR) derivation}\label{app:SNRnoise}
In this section, we derive the expression of the SNR, defined as
\begin{equation}
    \mathrm{SNR} = \frac{ \overline{ \tilde S_{ij} }^2 } { \mathrm{Var}(\tilde S_{ij} ) }\;,
\end{equation}
where $i\neq j$ and $\tilde S_{ij}$ is the signal measured by the balanced detector.
Differently than Sec.~\ref{app:variance_derivation}, we consider here the presence of noise $N_i$ (hence the different symbol $\tilde S_{ij}$), which is added to the received signal $\tilde Y_i = Y_i + N_i$.
Noise related to different signals are uncorrelated, i.e., $\overline{ N_i N_j^* } = 0$, with $i \neq j$, they are stationary and all characterized by zero mean value and an autocorrelation $\mathcal{R}_N(\tau) = \overline{ N_i(t)^* N_i(t+\tau) } = P_N \exp(-\Delta\nu_N |\tau|)$, where $\Delta\nu_N$ is the cut-off frequency of the transmission channel, and $P_N$ is the noise power.

The expectation value of the detected signal in the presence of noise is
\begin{equation}
    \overline{ \tilde S_{ij} } 
    = \int_{-T/2}^{T/2} \frac{2 ~ \mathrm{Re} \left( \overline{ \tilde Y_i \tilde Y_j^* } \right) } { T ({N_f}-1) } dt 
    = \int_{-T/2}^{T/2}\frac{2 ~ \mathrm{Re} \left( \overline{ Y_i Y_j^* } + \overline{ Y_i N_j^* } + \overline{ N_i Y_j^* } + \overline{ N_i N_j^* }  \right) } { T ({N_f}-1) } dt \;.
\end{equation}
Since the noise and the signals are uncorrelated ($ \overline{ Y_i N_j^* } = \overline{ N_i Y_j^* } = 0$) and we know that $\overline{ N_i N_j^* } = 0$, the expression simplifies to
\begin{equation}
    \overline{ \tilde S_{ij} } 
    = \frac{ 2 P_\mathrm{out} \mathrm{Re} (\gamma_{ij}) } { T ({N_f}-1) } \;,
\end{equation}
which is the same expression obtained without the noise [see Eq.~(\ref{eq:exp_value})].

As for the second moment of the detected signal, we have 
\begin{equation}
    \overline{\tilde S_{ij}^2} = \frac{2}{T^2 ({N_f}-1)^2}  \int_{T^2} ~ \mathrm{Re} \Big\{ \overbrace{ \overline{ \left[\tilde Y_i^*(t_1) \tilde Y_j(t_1) + c.c.\right] \tilde Y_i^*(t_2) \tilde Y_j(t_2) } }^{\tilde{\mathcal{I}}} \Big\} dt_1  dt_2\;.
\end{equation}
We can analyze the two terms of $\mathcal{I}$ separately:
\begin{equation}
    \tilde{\mathcal{I}} 
    = \overbrace{ \overline{ \tilde Y_i^*(t_1) \tilde Y_j(t_1) \tilde Y_i^*(t_2) \tilde Y_j(t_2) } }^{\eta_1} + \overbrace{ \overline{ \tilde Y_i(t_1) \tilde Y_j^*(t_1) \tilde Y_i^*(t_2) \tilde Y_j(t_2) } }^{\eta_2} \;.
\end{equation}
Of the $16$ terms obtained expanding the product in $\eta_1$, the only non-zero term is
\begin{equation}
    \eta_1 = \overline{ Y_i^*(t_1)Y_j(t_1)Y_i^*(t_2)Y_j(t_2) } = \sum_{n,m,p,q}^N t_{in}^* t_{jm} t_{ip}^* t_{jq} \Gamma_{nmpq}(t_1-t_2)\;,
\end{equation}
where we used the definition of the linear port $Y_i = \sum_{n=1}^{N_f} t_{in} X_n$ and the definition of $\Gamma_{nmpq}(t_1-t_2)$ given in Eq.~(\ref{eq:definitionGamma}).
Using the expression derived in Eq.~(\ref{eq:second_moment_Gamma_to_g1}) we get
\begin{equation}
    \eta_1 =  P_\mathrm{out}^2 \gamma_{ij}^2 \left(1 + |g^{(1)}(\tau)|^2\right) \;.
\end{equation}
For $\eta_2$, instead, more terms are non-zero:
\begin{equation}
\begin{aligned}
    \eta_2 &= \sum_{n,m,p,q}^{N_f} t_{jn}^* t_{im} t_{ip}^* t_{jq} \Gamma_{nmpq}(t_1-t_2) 
    + |\mathcal{R}_N(\tau)|^2 + \\
    &\qquad+    \sum_{n,m}^{N_f} t_{in} t_{im}^* \overline{X_n(t_1) X_m^*(t_2)} \mathcal{R}(-\tau)
    + \sum_{n,m}^{N_f} t_{jn}^* t_{jm} \overline{X_n^*(t_1) X_m(t_2)} \mathcal{R}(\tau)\;,
\end{aligned}
\end{equation}
where $\tau = t_1 - t_2$.
Using again Eq.~(\ref{eq:second_moment_Gamma_to_g1}) and considering that $\gamma_{ii} = \gamma_{jj} = 1$ we obtain
\begin{equation}
    \eta_2 = P_\mathrm{out}^2 \left[|\gamma_{ij}|^2 + |g^{(1)}(\tau)|^2\right]
    + |\mathcal{R}_N(\tau)|^2 + P_\mathrm{out}  \mathrm{Re}\left[g^{(1)}(\tau) \mathcal{R}_N(-\tau)\right] \;.
\end{equation}
To solve the integral we need the following relations:
\begin{equation}
    \frac{1}{T} \int_{-T/2}^{T/2} ~ |g^{(1)}(\tau)|^2 d\tau
   = \frac{1 - \exp(- \Delta\nu T)}{\Delta\nu T} \approx \frac{1}{\Delta\nu T}\;,
\end{equation}
\begin{equation}
    \frac{1}{T} \int_{-T/2}^{T/2} ~ |\mathcal{R}_N(\tau)|^2 d\tau
    =  \frac{P_N^2 \left[ 1 - \exp(- \Delta\nu_N T) \right] }{\Delta\nu_N T}
    \approx \frac{P_N^2}{\Delta\nu_N T}\;,
\end{equation}
\begin{equation}
    \frac{1}{T} \int_{-T/2}^{T/2} ~ \mathrm{Re}\left\{ g^{(1)}(\tau) \mathcal{R}_N(-\tau) \right\} d\tau
    \approx \frac{P_N (\Delta\nu + \Delta\nu_N) } { T \left[ (\Delta\nu + \Delta\nu_N)^2 + 4 \pi^2 \nu_0^2 \right] } \to 0 \;.
\end{equation}
The approximate result comes from considering $T \gg 1/\Delta\nu$, $T \gg 1/\Delta\nu_N$ and $\nu_0 \gg \Delta\nu, \Delta\nu_N$.
Therefore, going back to the expression of the second moment with the approximate results of the integrals we get
\begin{equation}
    \overline{ \tilde S_{ij}^2 }
    = \frac{2 P_\mathrm{out}^2}{(N_f -1)^2}  \left[ \mathrm{Re}\left(\gamma_{ij}^2\right)  + |\gamma_{ij}|^2  + \frac{1 + \mathrm{Re}\left(\gamma_{ij}^2\right)}{\Delta\nu T} + \frac{P_N^2}{P_\mathrm{out}^2 \Delta\nu_N T} \right]\;.
\end{equation}
Considering that $|\gamma_{ij}|^2 + \mathrm{Re}(\gamma_{ij}^2) = 2\mathrm{Re}(\gamma_{ij})^2$, we finally derive the variance of the detected signal:
\begin{equation}
    \mathrm{Var}(\tilde S_{ij})
    = \overline{ \tilde S_{ij}^2 } - \left(\overline{\tilde S_{ij}}\right)^2 
    = \frac{2 P_\mathrm{out}^2}{(N_f -1)^2}  \left[ 
    \frac{1 + \mathrm{Re}\left(\gamma_{ij}^2\right)}{\Delta\nu T} + \frac{P_N^2}{P_\mathrm{out}^2 \Delta\nu_N T} \right]\;.
\end{equation}
Finally, the expression of the SNR is the following
\begin{equation}
    \mathrm{SNR} = \frac{ \overline{ \tilde S_{ij} }^2 } { \mathrm{Var}( \tilde S_{ij} ) } 
    = \left[ \frac{1 + \mathrm{Re}(\gamma_{ij}^2)}{2 \mathrm{Re}(\gamma_{ij})^2} \frac{1}{\Delta\nu T } + \frac{P_N^2}{2 P_\mathrm{out}^2 \mathrm{Re}(\gamma_{ij})^2 } \frac{1}{\Delta\nu_N T} \right]^{-1} \;.
\end{equation}
In case we choose to consider only shot noise, the power spectral density for a balanced detector is $\mathcal{S}_{NN} = P_N^2/\Delta\nu_N = 2 h \nu_0 P_\mathrm{out}$.
Moreover, considering a transmission efficiency $P_\mathrm{out}/P_\mathrm{in} = \eta$, the final expression of the $\mathrm{SNR}$ is
\begin{equation}
    \mathrm{SNR} = \left[ \frac{ 1 + \mathrm{Re}(\gamma_{ij}^2) } { 2 \mathrm{Re}(\gamma_{ij})^2 ~ \Delta\nu T } + \frac{ h \nu_0 } { \eta P_\mathrm{in} \mathrm{Re}(\gamma_{ij})^2 ~ T } \right]^{-1} \;.
\end{equation}


\end{document}